\documentclass[preprint2]{aastex}

\received{18 Jan 2002}
\accepted{27 Sep 2002}
\journalid{}{}
\articleid{}{}

\slugcomment{}

\lefthead{}
\righthead{}

\begin{document}

\title{A Spectrophotometric Method to Determine the Inclination 
       of Class I Objects}

\author{Takeshi Nakazato\altaffilmark{1}, Taishi Nakamoto\altaffilmark{2} \\ 
and \\ Masayuki Umemura\altaffilmark{3}}
\affil{Center for Computational Physics, University of
  Tsukuba, Tsukuba 305-8577, Japan }

\altaffiltext{1}{e-mail: nakazato@rccp.tsukuba.ac.jp}
\altaffiltext{2}{e-mail: nakamoto@rccp.tsukuba.ac.jp}
\altaffiltext{3}{e-mail: umemura@rccp.tsukuba.ac.jp}

\begin{abstract}
A new method which enables us to estimate the inclination of Class I young 
stellar objects is proposed. 
Since Class I objects are not spherically symmetric, it is likely 
that the observed feature is sensitive to the inclination of the system. 
Thus, we construct 
a protostar model by carefully treating two-dimensional (2D) 
radiative transfer and radiative equilibrium. 
We show from the present 2D 
numerical simulations that the emergent luminosity $L_{\rm SED}$,
which is the frequency integration of spectral energy distribution (SED), 
depends strongly on the inclination of 
the system $i$, whereas the peak flux is insensitive to $i$.
Based on this result, we introduce a novel indicator $f_{\rm L}$, 
which is the ratio of 
$L_{\rm SED}$ to the peak flux, as a good measure for the inclination.
By using $f_{\rm L}$, we can determine the inclination regardless of the 
other physical parameters.  
The inclination would be determined by $f_{\rm L}$ 
within the accuracy of $\pm 5^{\circ}$, if the opening angle of 
bipolar outflows is specified by any other procedure.
Since this spectrophotometric method is easier than a geometrical method 
or a full SED fitting method, this method could be 
a powerful tool to investigate the feature of protostars statistically with 
observational data which will be provided by future missions, 
such as {\it SIRTF}, {\it ASTRO-F}, and {\it ALMA}.
\end{abstract}


\section{Introduction}

Low mass pre-main-sequence stars are classified into four categories, 
Class 0, I, II, and III, mainly according to the shape of their spectral 
energy distributions (SEDs) (Lada \& Wilking 1984; Adams, Lada, \& Shu 1987, 
Andr\'{e}, Ward-Thompson, \& Barsony 1993).
The shape of SED is strongly dependent on the structure surrounding 
a central star, such as an envelope, an outflow, a disk, 
and so on. 
For example, since Class 0 and Class I objects show a peak in their SEDs in a 
sub-millimeter range, they are thought to be surrounded by a cool massive 
envelope, which heavily obscures a hot central part of the object. 
It is considered generally that the order of the classification, 0, I, II, 
and III corresponds to an evolutional sequence of a low mass pre-main-sequence 
star. 
Class 0 and I objects are thought to be protostars, which are enshrouded by  
a thick envelope, and to be younger than T Tauri stars, which are classified 
as Class II and III objects. 

A protostar is not a spherically symmetric 
system as implied by the presence of a bipolar outflow. 
This asymmetry could come from a structure such as a central star with a disk 
system, which seems to be the progenitor of a planetary system. 
Thus, a careful scrutiny of the structure of a protostar system is 
important to understand the formation process of a star and/or 
planetary system. 
However, it is not easy to reveal the structure directly by imaging 
observations, 
because optical imaging is significantly subject 
to the optically thick envelope 
and also radio imaging is strongly limited by the resolution of the 
technique. 
In contrast, a spectrophotometric method using SEDs seems to be effective 
to investigate a large number of 
protostar systems (Kenyon, Calvet, \& Hartmann 1993; Men'shchikov, Henning, \& 
Fischer 1999; Nakamoto \& Kikuchi 1999).
For instance, Kenyon et al. (1993) modeled a protostar 
with incorporating spherically symmetric radiative transfer 
and simulated its SED. 
They fitted simulated SEDs with observed SEDs and inferred the structure of 
Class I objects.
The accuracy of the spectrophotometric method depends on 
the accuracy of the model of 
protostar, which is characterized by the density and temperature 
distributions in the envelope, 
the inclination of the system, and so forth.
In particular, 
the inclination of the system is an important piece of information to probe 
the inner structure of the protostar. 
It can affect the SED significantly through 2D behavior of radiative transfer.
Hence, the inclination is of essential importance in a realistic 
model of protostars. 
In other words, there may be a possibility to estimate the 
inclination using the SED. 
So far, 
the inclination of a protostar has been estimated by the geometrical feature 
of molecular outflow lobes.
However, the geometrical 
method requires high spatial resolution of observations and long 
integration time. 
Thus, it is not effective for a statistical study.
On the other hand, 
if we can estimate the inclination of the protostar solely by the SED, 
it would be quite effective to investigate the protostar structure. 

In this paper, we propose a new method to determine the inclination of 
Class I objects using their SEDs. 
We solve the radiative equilibrium for the 2D axisymmetric density 
distribution with incorporating 2D radiative transfer 
and simulate SEDs for the 2D protostar model. 
Then, it is shown that the ratio of the quantity $L_{\rm SED}$, 
which is the integration of the SED over frequencies, to 
the peak flux of the SED, $(\nu L_{\nu})_{\rm max}$, can be a good measure 
of the inclination. 
The accuracy and the applicability of the present method is discussed as well.

This paper is organized as follows.
In section 2 and 3, our model of a protostar and calculation procedure of the 
new method will be described. 
The new method will be introduced in section 4. 
The validity, reliability, and accuracy of the new method is also discussed 
in this section. 
The conclusions of this paper will be given in section 5.

\section{Model}

The present two-dimensional axisymmetric model for a Class I object 
consists of three components; a central star, a circumstellar disk, 
and an envelope surrounding the former two components.
The main opacity source is dust grains.
It is assumed that the dust grains are well mixed with hydrogen molecule
gas and the mass ratio to the hydrogen gas is $0.01$.
The opacity of dust grains as a function of the frequency is taken from 
Miyake \& Nakagawa (1993).
Physical parameters that characterize our model are summarized in 
Table \ref{param}.

The central star is only the energy source and radiative heating is considered.
Compressional heating, viscous heating, chemical heating, and any kind 
of heating sources other than the radiation from the central star are
ignored, since they are negligibly small in almost all the regions of interest.
The central star is assumed to have the surface temperature $T_{\ast}$ 
and emit the blackbody radiation with that temperature. 
The luminosity of the central star is given by $L_{\ast}$.
The mass of the central star is denoted by $M_{\ast}$.

Our model contains a circumstellar disk, although the presence of 
the disk has not been confirmed observationally in Class I objects.
Its inner and outer radii are assumed to be $0.1{\rm AU}$ and $100{\rm AU}$, 
respectively.
The surface density distribution of the disk is given by
$\Sigma( r ) = \Sigma_1 (r/1{\rm AU})^{-p}$, 
where $r$ is the radius in cylindrical coordinates from the central star,  
$\Sigma_1$ is surface density at $1{\rm AU}$, 
and $p$ is the index of the power, respectively.
We adopt $p=1.5$ which is a standard value for the 
minimum mass solar nebula model (Hayashi, Nakazawa, \& Nakagawa 1985).
The value of $p$ does not affect the SEDs significantly.
Vertical structure in the disk is determined by the hydrostatic equilibrium
between gravity by the central star and thermal pressure.

We assume the outer radius of the envelope to be $1,000{\rm AU}$.
Contributions to SED from the outside of that region are neglected.
This assumption seems appropriate, since the 
observed region of a Class I object 
is as small as a few thousand AU.
In the envelope, the density distribution is given by
\begin{equation}
   \rho( R ) = \rho_1 \left( \frac{R}{1{\rm AU}} \right)^{-q},
   \mbox{ for $1{\rm AU}<R<1,000{\rm AU}$, }
   \label{rho_env}
\end{equation}
where $R$ is distance from the central star in spherical coordinates, 
$\rho_1$ is mass density at $1{\rm AU}$, 
and $q$ is the index of the power, respectively.
Here, we adopt $q=1.5$ because it is the value expected for the 
free-falling envelope (Shu 1977).
Observationally, it is known that Class I objects usually have bipolar 
outflows, 
and it is suggested that the outflow excavates the envelope 
and affects the SED considerably. 
In order to take into account the effects of the bipolar outflow, 
we situate a cavity along the symmetric axis.
The cavity is assumed to be a conical shape with a half opening angle 
$\theta_{\rm bp}$, which is the angle between the symmetric axis and the 
boundary of the cavity. 
The density in the cavity is assumed to be 
0.01 times smaller than the value which is given by eq.(\ref{rho_env}).
Thus, the density in the envelope is given by
\begin{equation}
   \rho(R,\theta) = \left\{
                      \begin{array}{@{\,}ll}
                        \rho(R) & \mbox{ for $\theta>\theta_{\rm bp}$, } \\
                        0.01\times \rho(R) & 
                                  \mbox{ for $\theta<\theta_{\rm bp}$. } 
                      \end{array}
                    \right.
   \label{cavity}
\end{equation}

\section{2D Radiative Equilibrium Calculations}

\subsection{Numerical Method}

In the first step,
we obtain the temperature distribution in the model 
under the condition of the radiative equilibrium;
$\int_0^{\infty} \chi_{\nu}^{\rm abs} J_{\nu} d\nu
= \int_0^{\infty} \chi_{\nu}^{\rm abs} B_{\nu} d\nu$,
where $\chi_{\nu}^{\rm abs}$ is the absorption coefficient,
 $J_{\nu}$ is the mean intensity, 
and $B_{\nu}$ is the Planck function, respectively.
To calculate the radiative equilibrium, 
we solve 2D radiative transfer by a numerical scheme based on the
so-called Variable Eddington Factor method (Stone, Mihalas, \& Norman 1992; 
Kikuchi, Nakamoto, \& Ogochi 2002).
The primary difference of our algorithm from that by Stone et al. is that 
we treat a frequency dependence of radiative transfer accurately. 
In our method, the radiation field is separated into two parts; 
one is the direct component from the central star  
and the other is the diffuse component.
The direct component is calculated straightforwardly. 
The diffuse component is obtained by solving moment equations 
(Stone et al. 1992), 
which are extended to a non-gray version in our method, 
coupled with the energy equation. 
To close the moment equations,
the variable Eddington factor is introduced, 
which is determined by solving the radiative 
transfer equation taking the scattering effect into account.
We expand the scattering phase function by the Legendre function to the 
2nd order.
The set of equations are integrated with time. 
When the temporal change of the temperature distribution in the calculation 
region becomes small enough, we regard it as an equilibrium state.
In the second step, 
the SED of the object is simulated by the direct ray-tracing 
with the radiative equilibrium temperature distribution.

Our numerical code is based on cylindrical coordinates.
We use $50\times 50$ grids for $r$-$z$ coordinates in space,
and $50\times 100$ grids for azimuthal and zenith angles to express directions 
in which the intensity propagates.
Calculations with double number of grids in each dimension 
(16 times larger calculations) showed that the difference of results was 
within about 10\% in $f_{\rm L}$, which will be defined in eq.(\ref{fL}).
The frequency space is divided into 51 grids.

\subsection{SED calculation}
\begin{figure}
\epsscale{1}
\plotone{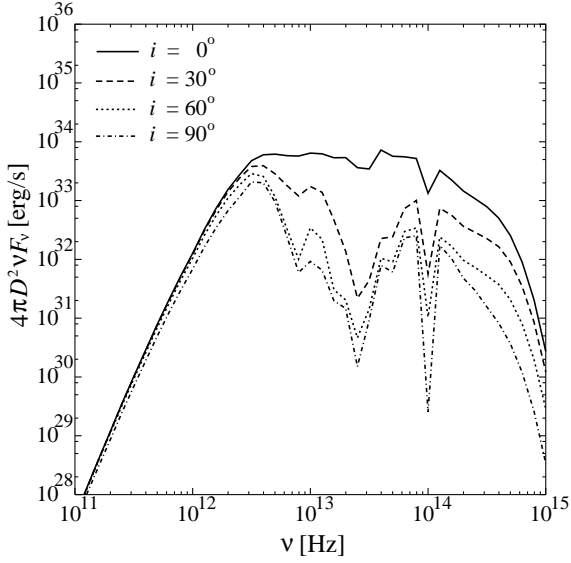}
\caption{
Emergent spectral energy distribution (SED) with the present protostar model.
Protostar model parameters are 
$L_{\ast}=1L_{\odot}$,
$\rho_1=10^{-13}\mbox{ g cm}^{-3}$,
$\Sigma_1=2000\mbox{ g cm}^{-2}$,
and $\theta_{\rm bp}=25^{\circ}$.
Four curves represent SEDs with the inclination
$i=0^{\circ}$ (solid curve),
$30^{\circ}$ (dashed curve),
$60^{\circ}$ (dotted curve),
and $90^{\circ}$(dot-dashed curve), respectively.
It is seen that the SED changes with the inclination $i$, even though 
the protostar structure does not change. 
The emergent luminosity $L_{\rm SED}$ also changes with the inclination. 
}
\label{sed}
\end{figure}

Simulated SEDs are illustrated in Fig. \ref{sed}.
It is seen that SEDs are sensitive to $i$, 
particularly in the frequency range 
higher than the peak frequency $\nu_{\rm peak}$.
This change reflects the asymmetry of the density and temperature distribution.
The optical depth along the line-of-sight 
from an observer to the central star, $\tau_{\rm c}$,  
is $\sim 1$ at $\nu_{\rm peak}$, while $\tau_{\rm c}>1$ for 
$\nu > \nu_{\rm peak}$ and $\tau_{\rm c}<1$ for $\nu < \nu_{\rm peak}$.
Thus, for $\nu<\nu_{\rm peak}$, the dependence on the inclination is within 
a factor of $2\sim 3$ and 
the emergent flux is the superposition of the thermal 
radiation from dust grains, which is determined by the temperature and 
the total amount of dust, regardless of density distributions. 
In contrast, for $\nu>\nu_{\rm peak}$, 
the anisotropy of flux is very large (2-4 orders of magnitude), 
and the emergent flux is dominated by 
the scattered radiation originating from the central star and inner disk,
which is strongly affected by the density distribution of the envelope. 
Therefore, the flux in the pole-on ($i=0^{\circ}$) direction is large, 
while the flux in the edge-on ($i=90^{\circ}$) direction is small.

\subsection{Comparison between Semi and Full 2D Calculations}

To see the importance of full 2D radiative transfer calculations, 
we performed approximate 
semi 2D calculations, which were carried out by Kenyon et al. (1993).
They derived the temperature distribution in the envelope 
from radiative equilibrium, 
but they assumed the spherically symmetric density distribution when they 
calculated the radiative equilibrium, so they obtained spherically symmetric 
temperature distribution. 
To do that, first the spherically averaged density profile 
from the 2D axially symmetric density distribution is obtained, 
and then the
1D spherically symmetric temperature profile is calculated. 
Using 
the spherical temperature distribution and the non-spherical 2D density 
distribution, the emergent SEDs from the object are obtained by ray-tracing .
Following their procedure, we also calculated the SED from the semi 2D 
calculations and compared the results with our full 2D calculations.

\begin{figure}
\epsscale{.52}
\plotone{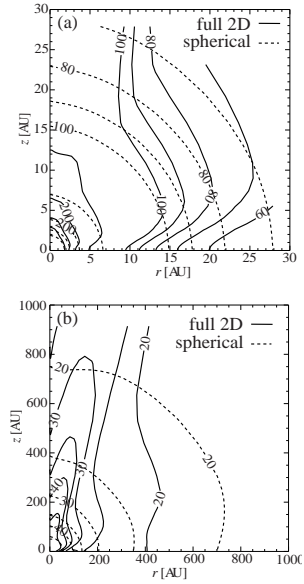}
\caption{
Temperature distributions obtained by the full 2D radiative equilibrium 
calculation (solid curve) and the spherically averaged 1D calculation 
(dotted curve) in regions (a) 30 AU $\times$ 30 AU and (b) 
1000 AU $\times$ 1000 AU.
Adopted parameters are 
$L_{\ast}=1.0L_{\odot}$, $\rho_1=10^{-13}\mbox{ g cm}^{-3}$, 
$\Sigma_1=2000\mbox{ g cm}^{-2}$, and $\theta_{\rm bp}=25^{\circ}$.
In the spherically averaged 1D calculation, density distribution is 
spherically averaged first, and then, the radiative equilibrium is solved 
with the 1D spherically symmetric radiative transfer calculation.
}
\label{temp}
\end{figure}

In Fig. \ref{temp}, we compare two temperature distributions derived from two 
procedures.
The temperature distribution obtained by the full 2D calculations is 
not spherically symmetric at all. 
The most noticeable difference between two temperatures is found in 
the disk and the outflow regions.  
The temperature in the disk in the full 2D calculation is much
lower than that of the spherical temperature, because the disk has higher 
density than the spherically averaged density.
We also find that the temperature in a shaded region, 
which is behind the disk with respect to
the central star, is lower than the spherical temperature, though 
the density in that region is not so different from spherically averaged one.
This is because that the total amount of radiation from the central star 
to the region is reduced by the efficient disk absorption. 
In contrast, the temperature in the outflow region in the full 2D 
calculation is higher than spherical
one, because the density in the region is lower than the 
spherically averaged one.
Consequently, it is obvious that the temperature distribution obtained by 
the full 2D radiative equilibrium calculation is quite different from 
spherically symmetric distribution.

\begin{figure}
\epsscale{.9}
\plotone{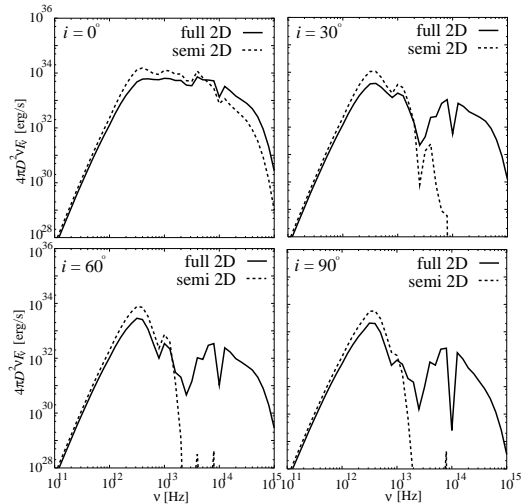}
\caption{
Spectral energy distributions based on temperature distributions 
obtained by two different methods 
shown in Fig. \ref{temp} for different inclination angles. 
SEDs from the non-spherical temperature distribution (obtained by the full 
2D radiative equilibrium calculation) are shown by solid curves, and 
those from the spherical temperature are shown by dotted curves.
}
\label{1d2d}
\end{figure}

Emergent SEDs based on 
the temperature distributions displayed in Fig. \ref{temp} 
are shown in Fig. \ref{1d2d}.
For each inclination, it is seen that (1) the flux derived from the full 2D 
calculations in a frequency range from  $10^{12}\mbox{ Hz}$ to 
$10^{13}\mbox{ Hz}$ is lower than that of the semi 2D calculations, and 
(2) in a range from  $10^{13}\mbox{ Hz}$ to $10^{15}\mbox{ Hz}$, 
the full 2D calculation flux is higher than the semi 2D 
calculation one.
The first difference is due to the decrease of the envelope temperature 
compared to the spherical temperature. 
Since the optical depth for the radiation at this frequency from the 
central star to the observer is around unity, the flux in this range is 
expected to be in proportion to the dust temperature in the envelope. 
On the other hand, the 
second difference is caused by the different mean intensity 
 $J_{\nu}$ especially in the outflow region, because flux in this frequency 
range is mainly determined by the scattering, which is  
evaluated by $\chi_{\nu}^{\rm sca} J_{\nu}$, 
where $\chi_{\nu}^{\rm sca}$ is scattering coefficient, 
and is very anisotropic in the non-spherical density distribution cases.
We can see from Figs. \ref{temp} and \ref{1d2d} that the full 2D radiative 
equilibrium calculations are indispensable to study the detailed structure 
of Class 0/I objects using their SEDs and/or images.

\section{Estimation of Inclination Angle}

\subsection{A New Inclination Indicator: $f_{\rm L}$}

\begin{figure}
\epsscale{.8}
\plotone{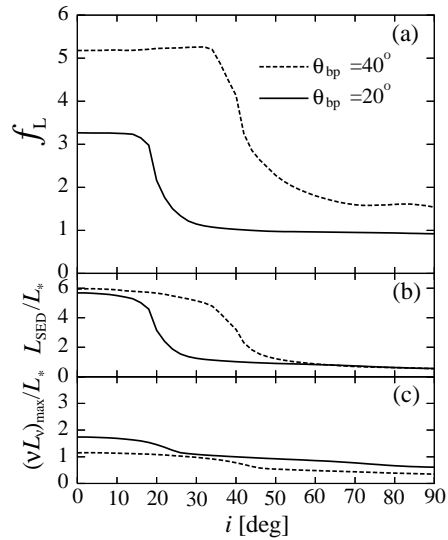}
\caption{
Three observables obtained directly 
from SED are plotted against the inclination. 
In each panel, two different curves correspond to SEDs with 
different opening angles in the outflow region, 
$\theta_{\rm bp}=20^{\circ}$ (solid curve), and 
$40^{\circ}$ (dotted curve), respectively: 
(a) the observable inclination indicator 
$f_{\rm L}$, which is the ratio of $L_{\rm SED}$ to 
$(\nu L_{\nu})_{\rm max}$, 
(b) the emergent luminosity $L_{\rm SED}$, and 
(c) the peak flux $(\nu L_{\nu})_{\rm max}$.
It is seen that $f_{\rm L}$ changes abruptly around $i=\theta_{\rm bp}$. 
If $f_{\rm L}$ is obtained from the SED, the inclination $i$ can be estimated 
for $i \ge \theta_{\rm bp}$ by using this figure.
}
\label{f_L}
\end{figure}

The inclination dependence of SED provides a tool to infer 
the inclination itself.
We introduce the emergent luminosity $L_{\rm SED}$, which is defined by 
\begin{equation}
   L_{\rm SED} = 4 \pi D^2 \int_0^{\infty} (\nu F_{\nu}) d\log\nu,
\label{LSED}
\end{equation}
as a tracer for the change of SED, where $D$ is the 
distance to the target object and $F_{\nu}$ is the observed flux 
at frequency $\nu$.
Also, using the peak flux 
$(\nu L_{\nu})_{\rm max}=(\nu\cdot 4\pi D^2 F_{\nu})_{\rm max}$ 
(flux at $\nu_{\rm peak}$), 
we define a ratio $f_{\rm L}$ by
\begin{equation}
   f_{\rm L} = \frac{L_{\rm SED}}{(\nu L_{\nu})_{\rm max}}.
\label{fL}
\end{equation}
(Note that $L_{\rm SED}$, $(\nu L_{\nu})_{\rm max}$, and $f_{\rm L}$ are 
all observable quantities.
We can obtain them from observed SEDs.)
We show the change of $f_{\rm L}$ against $i$ in Fig. \ref{f_L}.
Fig. \ref{f_L}(a) indicates that 
$i$ can be determined if $f_{\rm L}$ is evaluated from observed SED 
for the angle of $i \ge \theta_{\rm bp}$. 
But, when $i$ is in a range $i<\theta_{\rm bp}$, $f_{\rm L}$ 
is almost independent of 
$i$.
This means that SED does not change 
if we observe a protostar through the bipolar cavity.
For $i\sim 90^{\circ}$, $f_{\rm L}$ is also independent of $i$.
It corresponds to the case that we observe a protostar through the disk. 

The level of $f_{\rm L}$ depends on $\theta_{\rm bp}$.
It implies that $\theta_{\rm bp}$ is one of the important parameters 
to determine the anisotropy of a protostar system.
The ratio $f_{\rm L}$ increases for all $i$ as 
$\theta_{\rm bp}$ increases, and it is worth noting that 
there exist maximum and minimum of 
$f_{\rm L}$ for each $\theta_{\rm bp}$.

\begin{figure}
\epsscale{1}
\plotone{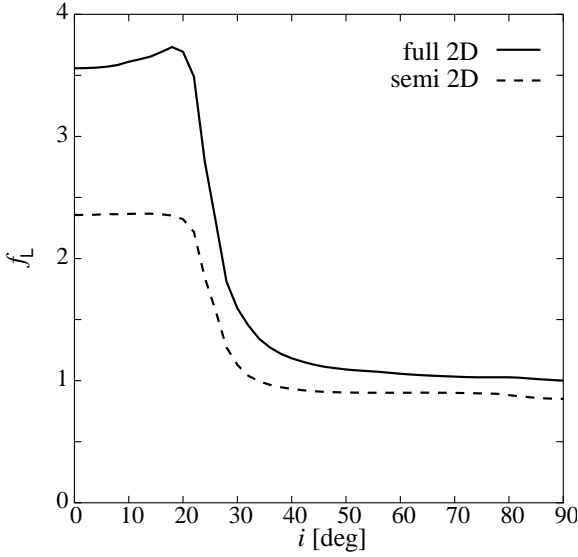}
\caption{The inclination indicator $f_{\rm L}$ 
based on the temperature distributions 
obtained by the semi 2D (dashed line) and 
the full 2D calculation (solid line). 
The values of $f_{\rm L}$ are different from each other 
especially for the small inclination.
}
\label{fL_1d2d}
\end{figure}

We calculated $f_{\rm L}$ for full 2D and semi 2D results shown in 
Figs. \ref{temp} and \ref{1d2d} to clarify the importance of the full 2D 
calculation. 
These results are shown in Fig. \ref{fL_1d2d}. 
In the semi 2D calculation, the value of $f_{\rm L}$ is lower than that of 
the full 2D calculation, especially for the  small inclination. 
It is because that the peak fluxes $(\nu L_{\nu})_{\rm max}$ are different by 
a factor of $\sim 2$ while the difference of observed luminosities 
$L_{\rm SED}$ is so small.
For the large inclination, 
the differences of $(\nu L_{\nu})_{\rm max}$ and $L_{\rm SED}$
are canceled each other so that the values of $f_{\rm L}$ are almost the same.

\subsection{Test of the Method}

We test our new method with a real object TMC1A (IRAS 04365+2535).
TMC1A is a Class I object (e.g. Kenyon et al. 1993; Chandler et al. 1998) and 
has an outflow with the half opening angle $\theta_{\rm bp}\sim 20^{\circ}$ 
(Chandler et al. 1996; Gomez et al. 1997).
According to Motte et al. (2001), TMC1A has an extended envelope and 
is recognized as a `true protostar' 
not a `transition object' nor an `edge-on view of T Tauri star'.
Thus, we think that TMC1A is an archetype of Class I objects 
and an appropriate example to test our new method to assess the inclination. 
Observational data of TMC1A are taken from Myers et al. (1987), 
Kenyon et al. (1993), and Chandler et al. (1998).

Here, we adopt the new method to derive the inclination, and compare the estimate with the results by the full SED fitting.
First, we attempt to estimate the inclination of the object using our new 
method.
We evaluate first the observed luminosity $L_{\rm SED}$ 
by integrating the SED  
and obtain $L_{\rm SED}=2.8L_{\odot}$, which is consistent with 
the results obtained by Myers et al. (1987) ($L=2.4L_{\odot}$) and 
Kenyon \& Hartmann (1995) ($L=2.2L_{\odot}$).
We next calculate $f_{\rm L}$ from $L_{\rm SED}$ and 
$(\nu L_{\nu})_{\rm max}$, and obtain $f_{\rm L}=2.52$.
From Fig. \ref{f_L}(a), this value indicates that inclination angle 
is about $20^{\circ}$. 
This inclination is estimated with our new method 
only using the $L_{\rm SED}$ and the peak flux, 
without fitting all the physical parameters of the object.

\begin{figure}
\epsscale{1}
\plotone{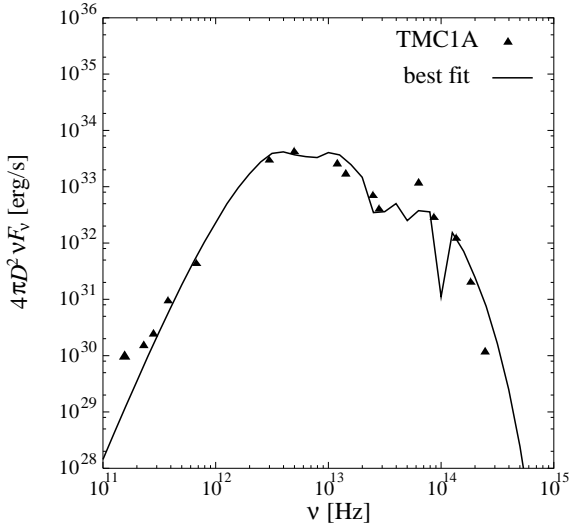}
\caption{
Best fitted SED with the full SED fitting procedure for TMC1A 
(IRAS 04365+2535). 
Adopted parameters are as follows: $L_{\ast}=1L_{\odot}$, 
$\rho_1=10^{-12.5}\mbox{ g cm}^{-3}$, $\Sigma_1=10^4\mbox{ g cm}^{-2}$, 
and $\theta_{\rm bp}=20^{\circ}$, and $i=22^{\circ}$.
Observational data of the object, plotted by filled triangles, are 
taken from Myers et al. (1987), Kenyon et al. (1993), 
and Chandler et al. (1998).
}
\label{TMC1A_SED}
\end{figure}

Next, we try to obtain all the physical parameters of the object by fitting 
the SED and estimate the inclination angle.
In practice, we fit the SED going through the following steps:
\begin{enumerate}
\item Set the outflow half opening angle $\theta_{\rm bp}=20^{\circ}$. 
\item Infer the luminosity of the central star $L_{\ast}$ from the peak flux 
of the SED.
\item Infer the circumstellar mass $M_{\rm env}+M_{\rm disk}$ 
($M_{\rm env}\propto \rho_1$, $M_{\rm disk}\propto \Sigma_1$) from the flux 
in a frequency range lower than the frequency of the peak flux 
$\nu_{\rm peak}$.
\item Infer the ratio of the envelope mass to the disk mass and inclination 
from the shape of the SED in a frequency range higher than $\nu_{\rm peak}$.
\end{enumerate}
The best fitted SED is displayed in Fig. \ref{TMC1A_SED} and adopted 
parameters are listed in Table \ref{TMC1A}.
It is seen that our model can reproduce the observational data very well. 
In this procedure, the inclination of the object is estimated to be 
$22^{\circ}$.

It is clear that the estimated values obtained by above two different methods 
agree quite well with each other.
Since the estimation based on the full SED fitting is considered to be more 
reliable, it seems that the new spectrophotometric method provides an 
effective tool to assess the inclination.

However, it should be noted that the inclination angle obtained above 
does not agree with the values obtained by Kenyon et al. (1993), which is 
$i\sim 60^{\circ}$, and 
Chandler et al. (1996), $i=40^{\circ}-70^{\circ}$. 
Such discrepancy could be attributed to the method used to infer 
the inclination.
Kenyon et al. (1993) used the spherically symmetric temperature distribution 
when they estimated the inclination with the SED.  
But the spherical temperature distribution can lead a significant difference 
in the resultant SED as shown in section 3.3. 
Chandler et al. (1996) estimated the inclination angle based on the shape of 
outflow lobe on the sky.
They first restricted the inclination angle to be less than $70^{\circ}$.
Then, they placed the lower limit of the inclination to be $40^{\circ}$, 
though the reason why they adopted the value was not given clearly.

\subsection{Effects of Other Parameters}

Here, we estimate 
uncertainties of the present method by changing parameters of the model.
As shown in Fig. \ref{f_L}(a), an uncertainty of $f_{\rm L}$ is large 
if the half opening angle $\theta_{\rm bp}$ is thoroughly unknown.
But, it is possible to pose a constraint for $\theta_{\rm bp}$ 
in the Class I phase 
by the observation of outflow lobes (Cabrit \& Bertout 1986).
For example, it is estimated that 
$\theta_{\rm bp}=13^{\circ}$ - $19^{\circ}$ 
for Class I object TMC1 (IRAS04381+2540) and 
$15^{\circ}$ - $21^{\circ}$ for TMC1A (IRAS04365+2535), 
respectively (Chandler et al. 1996).
There are also some other results for estimation of $\theta_{\rm bp}$: 
$\theta_{\rm bp}=22.5^{\circ}$ for L1448 (Bachiller et al. 1995) and 
B335 (Hirano et al. 1988; Chandler \& Sargent 1993), 
and $35^{\circ}$ for L1551 IRS5 (Andr\'{e} et al. 1990).
From these estimations, it seems possible to constrain 
$\theta_{\rm bp}$ to be $10^{\circ}$ - $40^{\circ}$.
This constraint would be justified by the fact that 
if $\theta_{\rm bp}$ becomes larger than this range, resultant SED exhibits  
double peaked feature in far-IR and near-IR to optical, but
these SEDs are not classified Class I category.
From Fig. \ref{fL}(a), a typical error for determination of $i$ 
is $\pm 10^{\circ}$ if we restrict ourselves to 
$\theta_{\rm bp}=10^{\circ}$ - $40^{\circ}$.

\begin{figure}
\epsscale{1}
\plotone{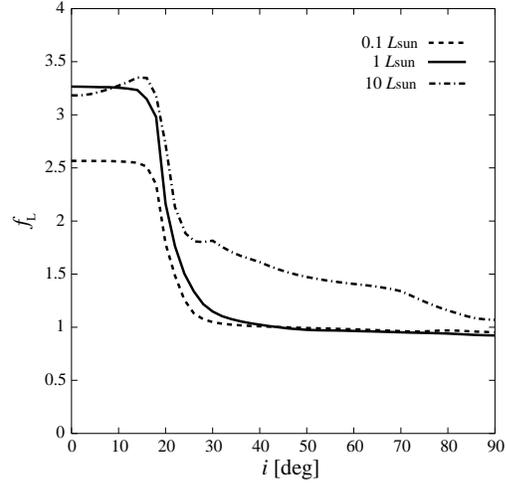}
\caption{
The inclination indicator, $f_{\rm L}$, vs the 
inclination, $i$, with three different luminosities.
When the luminosity is low, $f_{\rm L}$ for the small inclination is reduced, 
while if the luminosity is high, $f_{\rm L}$ for large inclination is raised. 
}
\label{lumi}
\end{figure}

Next, we examine to what extent the luminosity of the central star
$L_{\ast}$ affects the ratio $f_{\rm L}$.
Fig. \ref{lumi} shows  calculated $f_{\rm L}$ as functions of the inclination 
$i$ with three different luminosities, $0.1 L_{\odot}$, $1 L_{\odot}$, and 
$10 L_{\odot}$.
It is seen that when the luminosity of the central star is low 
($L_{\ast}=0.1 L_{\odot}$), $f_{\rm L}$ for the small inclinations decreases 
compared to the standard case ($L_{\ast}=1 L_{\odot}$), but $f_{\rm L}$ for
the large inclinations does not change. 
In contrast, when the luminosity is high ($L_{\ast}=10 L_{\odot}$), 
 $f_{\rm L}$ for the large inclinations becomes higher than the standard case, 
while $f_{\rm L}$ for the low inclinations remains the standard value. 
It is true that $f_{\rm L}$ is affected by the central star luminosity to 
some extent, but the change of the value in a range sensitive to the 
inclination is not large. 
Thus, when we estimate the inclination of an object using this new method and 
$f_{\rm L}$, the effect of the central star luminosity seems to be negligible.

\begin{figure}
\epsscale{1}
\plotone{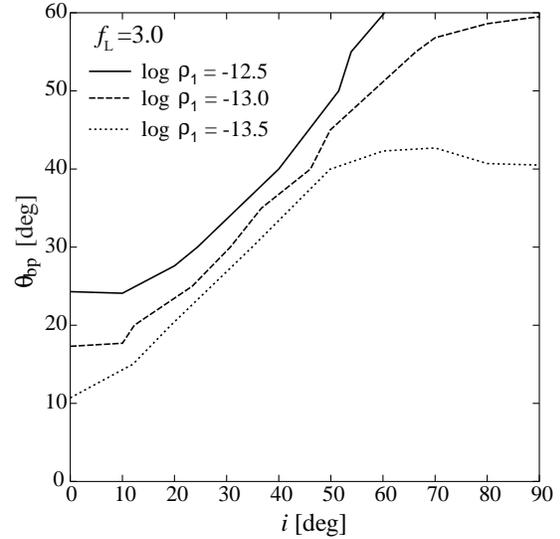}
\caption{The condition of $f_{\rm L}=3$ is plotted 
on $i$-$\theta_{\rm bp}$ plane.
Three curves correspond to three densities in the envelope at 1AU, 
$\rho_1=10^{-12.5}\mbox{g cm}^{-3}$ (solid curve), 
$\rho_1=10^{-13.0}\mbox{g cm}^{-3}$ (dashed curve), and 
$\rho_1=10^{-13.5}\mbox{g cm}^{-3}$ (dotted curve), respectively.
An error due to the change of $\rho_1$ is estimated to $\pm 5^{\circ}$ 
in the range of $\theta_{\rm bp}=0^{\circ}-40^{\circ}$. 
}
\label{rho}
\end{figure}

Also, $\rho_1$ could affect the results.
Fig. \ref{rho} shows the condition of $f_{\rm L}={\rm constant}$ 
for each $\rho_1$ in the $i-\theta_{\rm bp}$ plane.
The flux at $\nu>\nu_{\rm peak}$ decreases with increase of $\rho_1$, because 
$\tau_{\rm c}$ is proportional to $\rho_1$ and the flux in this frequency 
range is proportional to $\exp(-\tau_{\rm c})$.
In contrast, the flux at $\nu<\nu_{\rm peak}$ increases with increase of 
$\rho_1$ because the total amount of dust grains is increased.
Thus, $f_{\rm L}$ tends to decrease with increase of $\rho_1$, so that  
the line of $f_{\rm L}=\mbox{constant}$ is shifted above 
with increasing $\rho_1$ in the $i-\theta_{\rm bp}$ plane.
For $\theta_{\rm bp}=10^{\circ}-40^{\circ}$, the inclination $i$ 
changes by $\sim 10^{\circ}$ when $\rho_1$ changes by an order.

\begin{figure}
\epsscale{1}
\plotone{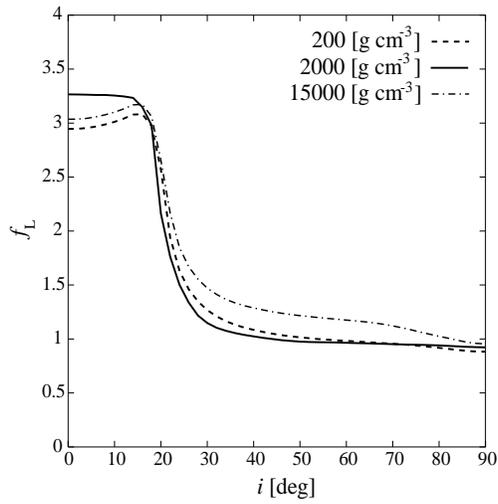}
\caption{
The inclination indicator, $f_{\rm L}$, vs the 
inclination, $i$, with three different disk masses. 
The disk mass does not affect the ratio $f_{\rm L}$ significantly.
}
\label{disk}
\end{figure}

By contrast, $\Sigma_1$ does not influence $f_{\rm L}$.
Fig. \ref{disk} displays calculated $f_{\rm L}$ with different $\Sigma_1$. 
It is easily seen that $\Sigma_1$ does not change $f_{\rm L}$ significantly. 
This is because the circumstellar disk contribute little 
to the column density along the line of sight, on which $f_{\rm L}$ strongly 
depends.

Therefore, we conclude that the primary uncertainty of our method 
comes from the information of $\theta_{\rm bp}$.
If $\theta_{\rm bp}$ can be estimated by any other means, 
then the error of inclination obtained by our method 
becomes roughly $\pm 5^{\circ}$.
If $\theta_{\rm bp}$ is unknown, then the error 
may be roughly $\pm 15^{\circ}$.  

\section{Conclusions}

Using 2D radiation transfer calculations, we have obtained the radiation 
fields for 2D axisymmetric protostar model, consistent with 
the central star, the circumstellar disk, and the envelope, 
and derived the SEDs of Class I objects.
We have found that the ratio $f_{\rm L}$ between the emergent luminosity, 
$L_{\rm SED}$, and the peak flux in the SED, $(\nu L_{\nu})_{\rm max}$, is a 
good indicator of the inclination angle of the object, because $L_{\rm SED}$ 
is sensitive to the inclination, while the peak flux is insensitive.
Through the test with real data and the analysis on the other physical 
parameters, it has been shown that $f_{\rm L}$ is a robust tool to assess 
the inclination of a Class I object.
Hence, $f_{\rm L}$ can provide a new spectrophotometric method 
to estimate the inclination angle of a Class I object. 
It is beneficial that, 
in this new method, only the luminosity $L_{\rm SED}$ and the peak flux 
$(\nu L_{\nu})_{\rm max}$ are needed to estimate the inclination. 
The typical error of the method is roughly $\pm 5^{\circ}$ if the half 
opening angle $\theta_{\rm bp}$ is known and roughly $\pm 15^{\circ}$ 
if it is unknown.   

The present method is applicable for a great deal of data provided by 
future missions of infrared, sub-millimeter, and millimeter observatories, 
e.g., {\it SIRTF}, {\it ASTRO-F}, and {\it ALMA}.
These missions are expected to reveal numerous protostar candidates. 
Then, it may be possible to carry out precise statistical 
study of protostars. 
The present spectrophotometric method could be a powerful tool 
in the statistical study.

\acknowledgments

We are grateful to Dr. N. Kikuchi for providing us the original 
numerical code. 
Numerical calculations were carried out with the facilities in 
Center for Computational Physics, University of Tsukuba. 
This work is supported in part by the Grant-in-Aid of the JSPS
(TN 12740118 and 10147105; MU 11640255).


\clearpage


\clearpage


\begin{table}
\begin{center}
\begin{tabular}{ccc}
\tableline
Symbol & Meanings & Parameter Range\\
\tableline
$L_{\ast}$ & luminosity of central star & $0.1-10L_{\odot}$\\
$T_{\ast}$ & temperature of central star & $4000{\rm K}$\\
$M_{\ast}$ & mass of central star & $0.5M_{\odot}$ \\
$\Sigma_1$ & surface density of disk at $1{\rm AU}$ 
& {$200-15000\mbox{ g cm}^{-2}$\,\tablenotemark{\it a}} \\
$p$ & power law index of disk & $1.5$\\
$\rho_1$ & density of envelope at $1{\rm AU}$ 
& {$10^{-13.5}$-$10^{-12.5} \mbox{ g cm}^{-3}$\,\tablenotemark{\it b}} \\
$q$ & power law index of envelope & $1.5$\\
$\theta_{\rm bp}$ & opening angle of bipolar cavity 
& $0^{\circ}$-$55^{\circ}$ \\
$i$ & inclination angle & $0^{\circ}$-$90^{\circ}$ \\ 
\tableline
\end{tabular}
\end{center}
\tablenotetext{\it a}
{This corresponds to the disk total mass $\sim0.03M_{\odot}$ assuming  
dust-to-gas mass ratio of $0.01$.}
\tablenotetext{\it b}{ $\rho_1= 10^{-13} \mbox{ g cm}^{-3}$ 
corresponds to the envelope total mass $\sim0.03M_{\odot}$ 
assuming dust-to-gas mass ratio of $0.01$ and $\theta_{\rm bp}=0^{\circ}$}.
\tablenum{1}
\caption{Parameters for the present protostar model.}
\label{param}
\end{table}

\clearpage

\setcounter{table}{1}
\begin{table}
\begin{center}
\begin{tabular}{|c||c|c|c|c|c||c|c|c|}
\tableline
Object & $L_{\ast}$ [$L_{\odot}$] & $\Sigma_1$ [g ${\rm cm}^{-2}$] &
$\rho_1$ [g ${\rm cm}^{-3}$] & $\theta_{\rm bp}$ & $i$ & 
$i(f_{\rm L})$ & KCH & Outflow \\
\tableline
TMC1A & 1.0 & $10^4$ & $10^{-12.5}$ & $20^{\circ}$ & $22^{\circ}$ & 
$20^{\circ}$ & $60^{\circ}$ & $40^{\circ}-70^{\circ}$\\
\tableline
\end{tabular}
\end{center}
\caption{Best fit parameters for TMC1A (IRAS 04365+2535).
From second to fifth columns show best fit parameters in our model for TMC1A. 
The sixth column shows the inclination angle estimated from $f_{\rm L}$. 
Seventh and eighth columns show the inclinations estimated from 
other work by Kenyon et al. (1993; marked `KCH'), who carried out full SED 
fitting using semi 2D protostar model,  
and Chandler et al. (1996; marked `Outflow'), who infered the inclination 
from the shape of outflow lobes on the sky.
}
\label{TMC1A}
\end{table}

\clearpage


\end{document}